\documentclass{elsart}
\usepackage{graphicx}
\usepackage{epsfig}
\usepackage{amssymb}
\usepackage{dcolumn}
\usepackage{bm}

\include{epsfig}

\journal{Physics Letter B}

\begin{document}

\begin{frontmatter}

\title{Separating hard and soft scales in hard processes in a QCD plasma\thanksref{energy}}\thanks[energy]{This work is supported in part by the US Department of Energy}

\author{A.H. Mueller}
\address{Department of Physics, Columbia University, New York, NY, 10027, USA}
\ead{amh@phys.columbia.edu}

\begin{abstract}
We present a picture of hard processes in a hot plasma in terms of the hard scale part of the process, where perturbative QCD should be applicable, and the soft scale part of the process, where we look to the AdS/CFT correspondence for guidance to possible strong effective coupling phenomena. In particular we estimate $\hat{q}$, the transport coefficient, supposing that at soft scales partons in the plasma all cascade to small-$x$-values as indicated by strong coupling SYM theory. 
\end{abstract}


\end{frontmatter}

\section{Introduction}
In perturbative QCD the transverse momentum broadening and energy loss of high energy jets in hot or cold matter \cite{Baier:1996sk,Baier:1998kq,Zakharov:1996fv,Wiedemann:2000za,Gyulassy:2000er} is governed by the transport coefficient, $\hat{q}$. That is 
\begin{equation}
  \frac{dp_{\perp}^2}{dt}=\hat{q} \label{qhat}
\end{equation}
and 
\begin{equation}
-\frac{dE}{dt}=\frac{\alpha_s N_c }{4}\hat{q}L \label{eloss}
\end{equation}
where Eq.~(\ref{eloss}) refers to the average energy loss of a high energy jet produced in the medium and then traversing a length $L$ of that medium. Traditional estimates\cite{Baier:1996sk} are that $\hat{q}$ is about $(0.5-1)\mbox{$GeV^2/fm$}$ for hot matter and $(0.02-0.04)\mbox{$GeV^2/fm$}$ for cold matter, for quark jets and a factor $\frac{N_c}{C_F}=\frac{9}{4}$ larger for gluon jets. It has been suggested that RHIC data favor a considerably larger value of $\hat{q}$ \cite{Eskola:2004cr,Dainese:2004te} and doubts have been cast on a purely perturbative picture for $\hat{q}$. In the context of the trailing string \cite{Herzog:2006gh,Gubser:2006bz} picture in the AdS/CFT correspondence to SYM theory, it is natural to have heavy quark energy loss and $p_{\perp}-$broadening which are considerably larger than that given in perturbative QCD. Also, in this picture, the concept of a local transport coefficient which governs these quantities seems to be lost; Eq.~(\ref{qhat}) and Eq.~(\ref{eloss})  are no longer valid, but are replaced by equations which are more nonlocal in the plasma\cite{Dominguez:2008vd}. 

On the other hand, one clearly can not trust a strong coupling SYM theory calculation for the energy loss of light quark and gluon jets in hot matter since genuine jets, collimated jets of energy, do not exist in such theories\cite{Hatta:2008tx,Hatta:2008tn}. What we propose here is to separate scales into hard and soft momentum regions\cite{Liu:2006he}. In the hard momentum region, where the jets are produced and where their evolution is on a hardness scale much greater than the temperature, $T$, we will trust perturbative QCD. Below some scale, $Q_{0}$, we suppose that perturbation theory may not be a good guide and we look to the strong coupling SYM theory-AdS/CFT correspondence for guidance. 

We shall first review the QCD calculation of $\hat{q}$ and then, after isolating the soft scale contributions to $\hat{q}$, we shall propose a picture of $\hat{q}$ where the soft scale contributions are estimated using a strong coupling SYM theory inspired picture. 

\section{Calculating $\hat{q}$ in perturbation theory and beyond}
In using Eq.~(\ref{qhat}) and Eq.~(\ref{eloss}) in hot matter, one can use perturbation theory to find, for gluon jets, 
\begin{equation}
\hat{q}_{g}=\frac{4\alpha N_c \pi^{2}}{N_{c}^2-1}\left(N_q xG_q+N_g x G_g\right), \label{qgluon}
\end{equation}
where $N_q$ and $N_g$ are the number densities of thermal quarks (and antiquarks) and thermal gluons in the plasma and $xG_q$  is the gluon distribution at the scale relevant for $\hat{q}$, coming from a single thermal quark and $xG_g$ is a similar quantity for a gluon. Using ideal gas distributions
\begin{eqnarray}
N_g &=&\frac{2(N_c^2-1)}{\pi^2}\zeta(3) T^3, \label{ng}\\
N_q &=&\frac{3N_c}{\pi^2}\zeta(3) T^3, \label{nq}
\end{eqnarray}
and 
\begin{equation}
 xG_q(x,Q^2)=\frac{C_F}{N_c} x G_g(x,Q^2), \label{xgluon}
\end{equation}
where $Q^2$ is the hard scale at which $\hat{q}$ is used, one has 
\begin{equation}
 \hat{q}_{g}=8\alpha N_c \zeta(3) T^3 xG_g\left(1+\frac{3}{4N_c}\right), \label{qgn}
\end{equation}
If $\alpha N_c \simeq 1$, $T=1/4 \mbox{GeV}$ and $xG_g\simeq 1$ in Eq.~(\ref{qgn}), one finds 
\begin{equation}
 \hat{q}_{g}=1 \mbox{$GeV^2/fm$}. \label{qgn2}
\end{equation}

The weak part of the above discussion is the use of ideal gas number densities in Eq.~(\ref{ng}) and Eq.~(\ref{nq}) and in the part of the evolution of $xG_g$  and $xG_q$ which extends to low scales. One can more confidently write
\begin{equation}
\hat{q}_{g}=\frac{4\alpha N_c \pi^{2}}{N_{c}^2-1} x g(x,Q^2), \label{qgluon2}
\end{equation}
where $xg$ in Eq.~(\ref{qgluon2}) is the gluon distribution per unit volume of the plasma. Eq.~(\ref{qgluon2}) follows from using perturbation theory at the scale $Q^2$ and this should surely be a good procedure so long as $Q^2$ is large. (In using $\hat{q}$  in Eq.~(\ref{qhat}) and Eq.~(\ref{eloss}) $Q$ should be taken to be the saturation momentum of the length of material through which the jet passes or, equivalently, the typical change in transverse momentum that jet experiences in passing through the matter.) The fact that $\hat{q}$ is meaningful and Eq.~(\ref{qhat}) and Eq.~(\ref{eloss}) apply follows from the hard part of the process alone with no assumption on the dynamics relevant at the scale $T$. 

\section{Going beyond perturbation theory: a QCD-SYM theory inspired picture}
In the perturbative QCD picture presented  above $xg$ is determined using lowest order evolution from the thermal scale $T$  to the hard scale $Q$. The parts of the evolution near $Q$ should be reliable while the parts of the evolution near the scale $T$ are, perhaps, suspect. In order to get a $\hat{q}$ larger than that coming from perturbation theory one would need a faster evolution at the soft momentum scales. Is there a physical reason why the actual evolution may be stronger than that indicated in perturbation theory? Strong coupling SYM theory points to a possible reason. In general as the coupling gets stronger, evolution becomes faster as more partons go to smaller x than at weak coupling. Strong coupling SYM theory, as evaluated using the AdS/CFT correspondence, is an extreme example of this. The structure function , $F_{2}(x,Q^2)$, of a dilaton\cite{Polchinski:2002jw}, or of the SYM plasma\cite{Hatta:2007cs} is very close to zero for $x>x_s(Q^2)$  because essentially $\underline{\mbox{all}}$ of the partons in this $x$-regime have disappeared by branching to lower values of $x$. This is the idea we shall take over, namely, that the effective coupling in the soft momentum regime may be strong enough that essentially all partons cascade to small values of $x$. As in strong coupling SYM theory we shall assume that the cascading stops when occupation numbers are on the order of one. For a given $Q_{0}^2$, this $x$-value is what we have called $x_s(Q_{0}^2)$. It is the $x$-value at which the saturation momentum is equal to $Q_{0}$\cite{Hatta:2007cs}. $Q_0$ will represent the transition point between weak coupling, for scales greater than $Q_0$, and our supposed strong coupling regime, for scales less than $Q_0$. The estimates we are about to present are admittedly crude and should be taken only as indicative of what might be happening. 

Now we are going to estimate the gluon density $xg$ in Eq.~(\ref{qgluon2})  compared to that of an ideal gas, the estimate depending on the parameter $Q_{0}$, the transition momentum between strong and weak coupling. To use the idea of saturation at the scale $Q_{0}^2$, we boost the plasma to a velocity $\tanh \eta$. Then for an ideal gas of gluons, the energy density in the boosted frame is 
\begin{equation}
E_{ideal}=N_g \frac{\pi^4}{30 \zeta(3)}T\cosh^2\eta, \label{ed}
\end{equation}
or
\begin{equation}
E_{ideal}\simeq 2.7N_g T\cosh^2\eta, \label{ed2}
\end{equation}
with $N_g$, given in Eq.~(\ref{ng}), the number density of gluons in the rest frame of a gluonic plasma assuming weak coupling. (We suppress quarks in this simple estimate.) Now in terms of the partons of the boosted plasma the energy density of a general plasma is given in terms of the gluon density $xg$ as 
\begin{equation}
E\simeq xg \cosh \eta Q_{0} \label{eq0}
\end{equation}
where $\cosh \eta$ is now chosen so that the partons longitudinal and transverse momentum are equal to $Q_{0}$ and the $x$-value of the gluons is $x_{s}(Q)$. There is a factor of $\cosh \eta$ in 
Eq.~(\ref{eq0}) because $xg$ is expressed as a density in the rest frame of the plasma. Equating Eq.~(\ref{ed2}) and Eq.~(\ref{eq0}), that is , supposing that the energy densities of an ideal gas and our plasma are the same gives
\begin{equation}
xg\simeq \frac{2.7T\cosh \eta}{Q_{0}}N_g, \label{xg}
\end{equation}
where the scale at which $xg$ is to be evaluated is $Q_{0}$. Before proceeding further to estimate $\cosh \eta$, let us interpret Eq.~(\ref{xg}). $2.7T$ is the typical thermal energy of a gluon in our ideal plasma in its rest system. $2.7T\cosh \eta$ is the energy of a thermal gluon in the boosted frame. The ratio $\frac{2.7T\cosh \eta}{Q_{0}}$ is then just $\frac{1}{x_s}$ with $x_s$ the value of $x$ to which the thermal gluon has cascaded in our strong coupling picture. Thus, we could equally well write 
\begin{equation}
xg\simeq \frac{1}{x_s(Q_{0})}N_g, \label{xg2}
\end{equation}
with the enhancement of gluons coming from our assumed cascading of all gluons to $x_s(Q_0)$. In the weak coupling limit there is little evolution and $xg$ would just be equal to $N_g$. (Recall that we took $xG\simeq1$ in going from Eq.~(\ref{qgn}) to Eq.~(\ref{qgn2}).)

Now we proceed to estimate $\cosh \eta /Q_{0}$ in Eq.~(\ref{xg}). Consider a part of the volume of the plasma which extends over $0\leq z\leq L$, $-\infty <x,y<\infty$ in the rest system of the plasma. $L$ is chosen so that this longitudinal region of the plasma contracts to a size $1/Q_{0}$ after boosting. That is 
\begin{equation}
L=\cosh \eta /Q_{0}. \label{lq}
\end{equation}
Now we write two equations for the longitudinal momentum per unit area of the slice of the plasma having $0\leq z \leq L$. In the boosted frame, one of these equation is 
\begin{equation}
\frac{dp_{z}}{d^2x_{\perp}}=\frac{2(N_c^2-1)\pi^2 T^4}{30}L \sinh \eta, \label{pa1}
\end{equation}
the momentum/area of an ideal gas of thermal gluons. Our second equation is in terms of partons,
\begin{equation}
\frac{dp_{z}}{d^2x_{\perp}}=\frac{2(N_c^2-1)}{(2\pi)^3}f(\pi Q_{0}^2) Q_{0} \label{pa2}
\end{equation}
with $f$ the quantum occupancy of the gluons. The $\pi Q_{0}^2$ in Eq.~(\ref{pa2}) is the transverse momentum phase space while the final $Q_{0}$ is the longitudinal momentum of the partons whose value matches the contracted size of our slab of matter as given by Eq.~(\ref{lq}). Comparing Eq.~(\ref{pa1}) and Eq.~(\ref{pa2}), after using Eq.~(\ref{lq}), gives (taking $\cosh \eta \simeq \sinh \eta$)
\begin{equation}
\cosh \eta =\sqrt{\frac{15}{4}f}\left(\frac{Q_{0}}{\pi T}\right)^2.\label{cosh1}
\end{equation}
Now taking $f\simeq 1$ gives
\begin{equation}
\cosh \eta \simeq 2\left(\frac{Q_{0}}{\pi T}\right)^2.\label{cosh2}
\end{equation}
Using Eq.~(\ref{cosh2}) in Eq.~(\ref{xg}) gives
\begin{equation}
xg\simeq \frac{Q_{0}}{2T}N_g.\label{xg3}
\end{equation}
Thus our estimate gives an enhancement of $\frac{Q_{0}}{2T}$ for the partonic gluon density at scale $Q_{0}$ and an identical factor for the enhancement to $\hat{q}$ as given in Eq.~(\ref{qgluon}), in the approximation where $xG_g \simeq 1$ and quarks are neglected. While we do not take the precise value we have found in Eq.~(\ref{xg3}) as definitive, we feel the mechanism we have used, a strong cascading (evolution) of partons from the large-$x$ to the small-$x$ region, should be natural $\underline{\mbox{if}}$ the effective coupling in the low momentum region is strong. It would be interesting to find more direct tests for a suppression of large-$x$ partons in the quark gluon plasma.

\end{document}